# Tunneling studies in a homogeneously disordered s-wave superconductor: NbN

S. P. Chockalingam[a1], Madhavi Chand[a], Anand Kamlapure[a], John Jesudasan[a], Archana Mishra[a,c], Vikram Tripathi[b], Pratap Raychaudhuri[a2]

[a]Department of Condensed Matter Physics and Materials Science, Tata Institute of Fundamental Research, Homi Bhabha Rd., Colaba, Mumbai 400005, India.

[b]Department of Theoretical Physics, Tata Institute of Fundamental Research, Homi Bhabha Rd., Colaba, Mumbai 400005, India.

[c]Institute Instrumentation Centre, Indian Institute of Technology Roorkee, Roorkee, Uttarakhand 247667, India.

***Abstract:*** We report the evolution of superconducting properties as a function of disorder in homogeneously disordered epitaxial NbN thin films grown on (100) MgO substrates, studied through a combination of electrical transport, Hall Effect and tunneling measurements. The thickness of all our films are >50nm much larger than the coherence length $\xi_0$~5nm. The effective disorder in different films encompasses a large range, with the Ioffe-Regel parameter varying in the range $k_F l$~1.38-8.77. Tunneling measurements on films with different disorder reveals that for films with large disorder the bulk superconducting transition temperature ($T_c$) is not associated with a vanishing of the superconducting energy gap, but rather a large broadening of the superconducting density of states. Our results provide strong evidence of the loss of superconductivity via phase-fluctuations in a disordered s-wave superconductor.

[1] Electronic-mail: chocka@tifr.res.in
[2] Electronic-mail: pratap@tifr.res.in

The effect of disorder on superconductivity[1,2,3,4] has been a topic of considerable experimental and theoretical interest for several decades. It was shown by Anderson[1] in the late 1950's that non-magnetic disorder is a *s*-wave superconductor should not significantly affect $T_c$ as long as the system remains a metal. However, later experimental investigations mostly on granular superconductors[2] showed that the $T_c$ gets gradually suppressed at the extreme disorder limit, eventually leading to a superconductor to insulator transition. It has been argued[5] that this regime, where the mean free path ($l$) becomes of the same order of magnitude as the inverse Fermi wave vector ($k_F$) is beyond the domain of validity of the early theories[1,2] which are applicable for $k_F l>>1$. Though a generally accepted physical picture of the evolution of superconducting properties in the presence of strong disorder has not yet emerged, it has been argued that both amplitude and phase fluctuations have to be incorporated to obtain a realistic picture of superconductivity in this regime. Numerical simulations indicate that a homogeneously disordered superconductor in the strong disorder limit breaks up into small superconducting domains whereby the loss of superconductivity is caused by a loss of the superfluid stiffness rather than the vanishing of the superconducting energy gap[5,6].

In this letter, we report the evolution of superconducting properties in homogeneously disordered epitaxial NbN films grown on (100) MgO substrates, through a combination of transport, Hall effect and tunneling studies. In a recent paper[7] we have reported that the carrier density ($n$) and $T_c$ in epitaxial NbN films is sensitive to the growth condition. As a result films with $k_F l$ varying from moderately clean to strongly disordered limit can be synthesized by controlling the deposition conditions during growth without destroying the crystallographic structure of NbN. These epitaxial films thus provide us with an ideal test bed for investigating the effect of homogeneous disorder in an s-wave superconductor. All the films reported in the present work have thickness >50nm which is much larger than the coherence length, $\xi_0$~5nm and are therefore in the three dimensional (3D) limit. The $k_F l$ values determined from the resistivity and Hall Effect vary in the range 1.38-8.77. Since $k_F l$~1 correspond to the Mott limit for metallic conductivity the level of disorder span from moderately clean to the very dirty limit. However, the sheet resistance of our most disordered film at 17K is ~2kΩ. We are therefore far from the Anderson superconductor-insulator insulator phase boundary, which has commonly been the focus of many works[2,8] on disordered superconductor. The tunneling density of states has been measured as a function temperature in the superconducting state on films with different disorder by fabricating on NbN/oxide/Ag planar tunnel junctions. The central result of this paper is that even far from the superconductor-insulator transition, in the presence of strong disorder, the bulk $T_c$ of an s-wave superconductor is not associated with a vanishing of the superconducting energy gap (Δ) but rather a large broadening of the quasiparticle density of states (DOS).

Epitaxial NbN samples were synthesized on (100) oriented single crystalline MgO substrates by dc reactive magnetron sputtering, by sputtering a Nb target in Ar/$N_2$ gas mixture. Films with different levels of disorder were synthesized by controlling the sputtering power during growth. All the films synthesized in this way were epitaxial in nature as revealed from X-ray ϕ–scan measurements. Details of synthesis and characterization of NbN thin films have been reported in ref. 7. Planar tunnel junctions were fabricated by first depositing a 300μm wide NbN strip and oxidizing it at $250^0$C in oxygen atmosphere for 2 hrs. Two tunnel junctions were fabricated on each device by subsequently evaporating 300μm cross strips of Ag. Under optimal growth conditions, this process resulted in highly reproducible tunnel junctions on almost every attempt, with high bias (V>>Δ/e) tunnel junction resistance varying between 2-10Ω. Tunnel junctions oxidized for longer time have higher junction resistance but poorer tunneling characteristics, as a result of the formation of defects in the oxide tunnel barrier. While all our tunnel junctions show excellent tunneling characteristics in the superconducting state, these could not be used for tunneling spectroscopy above $T_c$ due to the large resistance of the films compared to the tunnel junction. While fabricating the tunnel junctions the device was configured is such a way[9] that we could perform both I-V measurements on the tunnel junction as well as resistance versus temperature (R-T) measurements of the NbN film on which the tunnel junction was fabricated. The current versus voltage (I-V) characteristics of the tunnel junction was recorded at various temperatures using a Keithley 2001 source-meter. In all these measurement the voltage NbN electrode was connected to the virtual ground whereas the voltage was applied on the silver electrode. The conductance ($G(V) = \left.\frac{dI}{dV}\right|_V$) versus voltage (V) of the tunnel junction was obtained by numerically differentiating the I-V data. The $T_c$ of the film on which tunnel junction was fabricated, was independently measured from R-T measurements using standard 4-probe technique. Hall effect was measured on samples patterned in a Hall bar geometry. Runs of several samples showed that for identical growth conditions the $T_c$ of the films are within 15% of each other.

Figure 1(a) shows the variation of $T_c$ with $k_F l$ for NbN films with different levels of disorder. The values of $k_F l$ are determined from the $n$ extracted from Hall effect and the normal state resistivity $\rho_n$ (both measured at 18K) using free electron relations[10]. Figure 1(b) shows the variation of $\rho_n$ (measured at 17K) versus $k_F l$ for the same films. The inset of figures 1(a) and 1(b) show the variation of $T_c$ and normal state conductivity $\sigma_n(=1/\rho_n)$ with $n$. With decreasing $k_F l$ the $T_c$ decreases and $\rho_n$ increases. As a first approximation both the variation of $\rho_n$ and $T_c$ with disorder can be understood from the variation[7] in $n$. While the disorder tuning in these films is likely to be a result of Nb deficiencies in highly disordered films, the large change $n$ cannot be taken into account by chemical considerations alone[7]. This points towards an unusual localization of the electronic carriers with increase in disorder which will be dealt in a separate paper. Since we cannot perform Hall Effect on the same films on which the tunnel junction is fabricated, the $k_F l$ values of those films are estimated from the polynomial fit to the variation $T_c$ and $\rho_n$ with $k_F l$ and 1/ $k_F l$ respectively, shown in solid lines in Figure 1(a) and 1(b).

Figure 2(a-c) show the G(V) vs V spectra at different temperatures down to 2.2K for three films with different levels of disorder corresponding to (a) $T_c$=14.9K, $k_F l$~6, (b) $T_c$=9.5K, $k_F l$~2.3 and (c) $T_c$=7.7K, $k_F l$~1.4 respectively[11]. An asymmetry between positive and negative bias as well as a slope at high bias, similar to the tunneling spectra High $T_c$ cuprates[12,13,14], is observed for all the spectra. This asymmetry is larger for samples with lower $k_F l$. Before theoretically fitting the spectra, the spectra were symmetrised by subtracting a linear background passing through the origin[15]. The conductance spectra were fitted with the tunneling equation[16],

$$G(V) = \frac{dI}{dV}\bigg|_V = \frac{d}{dV}\left\{\int_0^{eV} N_s(E)N_n(E-eV)\{f(E)-f(E-eV)\}dE\right\}, \qquad (1)$$

where $N_s(E)$ and $N_n(E)=N_n$ are density of states of the superconducting and normal metal electrodes respectively and $f(E)$ is the Fermi-Dirac distribution function. For $N_s(E)$ we used the BCS DOS with an additional broadening parameter[17] Γ, so that, $N_s(E) = Re\left\{\frac{E-i\Gamma}{[(E-i\Gamma)^2-\Delta^2]^{1/2}}\right\}$. $\Gamma(=\frac{\hbar}{\tau})$ is formally introduced in this expression to take into account the effect of finite lifetime ($\tau$) of the superconducting quasiparticles. However, while fitting a spectrum, Γ phenomenologically incorporates all non-thermal sources of broadening in the BCS DOS. The symmetrised spectra along with the corresponding fits are shown in figure 2(d-f). The best fit values of $\Delta$ and $\Gamma$ at the lowest temperature ($T$~$2.2K$) is shown in Table 1. We observe an increase in $\Gamma_0/\Delta_0$ with increase in disorder similar to earlier observation in granular Al samples[2] and disordered TiN[8].

The central result of this letter, namely the temperature variation of $\Delta$ and $\Gamma$ obtained from the best fit values of the spectra are shown in figures 3(a-c). In the same graph we show the normalized R-T of the NbN films on which the tunnel junction is fabricated. For the film with least disorder ($k_F l$~6), $\Delta$ and $\Gamma$ follow the temperature variation expected for a strong coupling superconductor[18]. In this case $T_c$ is associated with a vanishing of the $\Delta$. For the films with large disorder on the other hand, $\Delta$ does not go to zero as the $T \rightarrow T_c$ but remains finite even for $T \approx T_c$. This is most clearly seen for the sample with $T_c$=7.7K, $k_F l$~1.4 (fig. 3(c)), where $\Delta$ decreases to only 60% of its low temperature value at 7.3K. On the other hand $\Gamma$ increases rapidly as the temperature approaches $T_c$. While it is not possible to reliably fit the spectra at temperatures very close to $T_c$, from the trend in variation of $\Delta$ and $\Gamma$ we see that $T_c$ phenomenologically corresponds to the point where $\Delta \approx \Gamma$.

In conventional mean field theories of s-wave superconductors, such as BCS theory or its strong coupling counterparts[19], the superconducting $T_c$ is naturally defined as the temperature where $\Delta$ goes to zero. This corresponds to the temperature where the amplitude of the superconducting order parameter vanishes. However, the zero resistance state in a superconductor is characterized not only by finite pairing amplitude but also by global phase coherence. Therefore, in principle bulk superconductivity can also be lost due to the loss of superfluid stiffness whereby phase fluctuations between different regions of the superconductor destroy the macroscopic zero resistance state even if the amplitude of the order parameter remains finite. This mechanism has been explored[20] in the context of Josephson junction (JJ) networks and granular superconducting films where the superconductor can be envisaged as a disordered network of Josephson junctions. Recent numerical simulations[5,21] on disordered 2-D superconductors as well as scanning tunneling spectroscopy experiments[8] on homogeneously disordered TiN films close to the superconductor-insulator phase boundary, reveal that the situation in a strongly disordered s-wave superconductor is similar to disordered JJ network. Under strong disorder the superconductor electronically segregates into superconducting domains separated by insulating regions. The bulk $T_c$ corresponds to the temperature where the global phase coherence[5] is lost between these islands[22]. In such a situation the amplitude of the order parameter (and hence $\Delta$), will not go to zero at $T \approx T_c$, but BCS DOS will broaden due to phase fluctuations. The non-vanishing $\Delta$ and the large increase in $\Gamma$ as $T \rightarrow T_c$ in strongly disordered ($k_F l \sim 1$) epitaxial NbN films[23] supports this scenario.

Extrapolating the $\Delta(T)$ versus $T$ data for the NbN film with strong disorder it is natural to assume that the gap in the DOS persists even in the normal state. This naturally points to the formation of a "pseudogap" state with a finite value of energy gap but no global phase coherence for $T > T_c$. In such a situation, the bulk $T_c$ observed from resistivity measurements is smaller than the mean field transition temperature ($T^*$) expected in the absence of phase fluctuations. While we cannot obtain spectroscopic information for $T > T_c$ due to low resistance of our tunnel junctions, an indirect evidence of this comes from the measurement $2\Delta_0/k_B T_c$ ($\Delta_0$ corresponds to the value of $\Delta$ at 2.2K) which is a measure of the electron-phonon coupling strength[16], within mean field theories of superconductivity. Figure 4 shows the variation of $2\Delta_0/k_B T_c$ as a function $T_c$ for six NbN films with different levels of disorder. For the least disordered film with $T_c$~14.9K, $2\Delta_0/k_BT_c$=4.36 as expected for a strong coupling superconductor. Since in our films $n$ decreases with increase in disorder, for films with lower $T_c$ the electron-phonon coupling strength is expected to decrease due to the decrease in density of states at Fermi level[7]. This trend is observed for films with $T_c$>9.5K. However in the highly disordered samples ($T_c$<9.5K) $2\Delta_0/k_BT_c$ shows an anomalous increase reaching a value of 4.43 for the film with $T_c$~7.7K. This increase signals a breakdown of the mean field scenario where the $T_c$ obtained from the onset of zero resistance is smaller than the mean field $T^*$. Therefore between $T_c$ and $T^*$ one would expect a state with a gap in the quasiparticle spectrum but no overall phase coherence. Such a state should in principle be detectable in scanning tunneling spectroscopy measurements at temperatures higher than $T_c$ where the tunneling resistance varies between few 100MΩ to a few GΩ and the contribution of the normal state resistance of the sample is insignificant.

While the loss of superfluid stiffness at $T_c$ seems the most natural explanation for the observed variation in $\Delta$ and $\Gamma$ in strongly disordered NbN, it is also worthwhile to explore whether other scenarios proposed in disordered superconductor could explain our data. It was suggested by Anderson, Muttalib and Ramakrishnan[24] that close to $k_Fl$~1 the loss of effective screening increases the electron-electron repulsion, leading to a decrease in $T_c$. Within McMillan theory of strong coupling superconductor this effect is leads to an increase of the effective Coulomb pseudo-potential $\mu^*$. While the $k_Fl$ of our films are in the range where this effect could manifest itself, the magnitude of this effect is difficult to estimate, since the critical resistivity at which superconductivity is destroyed is an adjustable parameter within this theory[25]. More importantly this mechanism gives a positive correction to the effective $\mu^*$ which does not explain the increase in $2\Delta/k_BT_c$ or our observation of finite $\Delta$ at $T\approx T_c$. Another model for disordered superconductor was suggested by Bergmann[26]. He argued that disorder causes an effective increase in the electron-phonon coupling strength due to phonon emission caused by inelastic scattering processes at impurity sites. It was qualitatively argued[27] that is strong coupling superconductors, this effect is compensated by "strong coupling" effects, which could in extreme situations cause a decrease in Δ and $T_c$. While this effect could in principle account for the increase in $2\Delta/k_BT_c$, it is not consistent with the non-vanishing of Δ at $T\approx T_c$.

In conclusion, we show that in three dimensional strongly disordered NbN epitaxial thin films, the superconducting energy gap does not vanish as $T\rightarrow T_c$ as expected for a conventional s-wave superconductor. Furthermore, $2\Delta_0/k_BT_c$ shows an anomalous increase with increasing disorder in this regime. These results are in good agreement with recent theoretical predictions on strongly disordered s-wave superconductors. Our results indicate that the DOS is likely to remain gapped even in the normal state at, resulting in a pseudogap state at temperatures $T>T_c$. In this context we would like to note than a recent theory[28] of building the superconducting state in the presence of strong disorder from fractal wavefunctions, also predicts the anomalous increase in $2\Delta_0/k_BT_c$ and the formation of a gapped normal state close to the superconductor-insulator phase boundary. However the theory outlined in ref. 28 is applicable close to the Anderson superconductor-insulator transition. All our films have sheet resistance <2kΩ and are therefore far from this boundary. The unique feature of our results is that even far from the Anderson supercondontuctor-insulator phase boundary we see that superconducting $T_c$ is strongly influenced by phase fluctuations when $k_Fl$ approaches 1. It would be instructive to obtain spectral information on the state at $T>T_c$, as well as its possible special variation using techniques such scanning tunneling spectroscopy. These measurements are currently underway are will be reported in a future paper.

*Acknowledgements:* We would like to thank T. V. Ramakrishnan and Nandini Trivedi for illuminating discussions, Vivas Bagwe for technical help and Sangita Bose for critically reading the manuscript.

**Table 1:** Superconducting transition temperature ($T_c$), normal state resistivity ($\rho_n$), superconducting energy gap ($\Delta_0$) and broadening parameter ($\Gamma_0$) measured at 2.2K for the three NbN films shown in Figure 3 (a)-(c).

| Sample | $T_c$ (K) | $\rho_n$ ($\mu\Omega$-m) | $\Delta_0$ (meV) | $\Gamma_0$ (meV) | $\Gamma_0/\Delta_0$ |
|---|---|---|---|---|---|
| (a) | 14.9K | 1.58 | 2.80 | 0.006 | 0.002 |
| (b) | 9.5K | 5.11 | 1.60 | 0.07 | 0.044 |
| (c) | 7.7K | 6.75 | 1.47 | 0.08 | 0.054 |

**Figure captions**

**Figure 1.** (a) Variation of $T_c$ with $k_F l$ in NbN films with different levels of disorder; the solid line is a polynomial fit to the data. The *inset* shows the variation of $T_c$ with the carrier density, $n$, extracted from Hall effect measurement at 17K. (b) Variation of $\rho_n$ with $k_F l$ for the same films; the solid line is a polynomial fit in powers of ($1/k_F l$). The *inset* shows the variation of $\sigma_n$ ($=1/\rho_n$) with $n$ (points) and the linear fit to the relation $\sigma_n \propto n$**.**

**Figure 2.** (a-c)*G(V)-V* spectra at different temperatures of NbN/oxide/Ag tunnel junction at different temperatures on films with different levels of disorder corresponding to (a) $T_c$=14.9K, $k_F l$~6, (b) $T_c$=9.5K, $k_F l$~2.3 and (c) $T_c$=7.7K, $k_F l$~1.4. Some temperature have been removed for clarity. (d-f) The same set of spectra corresponding to (a)-(c) after removing the asymmetric background (points) along with the corresponding theoretical fits. The fit is restricted to bias values equal to $2.5\Delta_0/e$ where $\Delta_0$ corresponds to the superconducting energy gap at the lowest temperature.

**Figure 3.** Temperature variation of $\Delta$ and $\Gamma$ and $\rho$ for three NbN films with different levels of disorder: (a) $T_c$=14.9K, $k_F l$~6, (b) $T_c$=9.5K, $k_F l$~2.3 and (c) $T_c$=7.7K, $k_F l$~1.4.

**Figure 4.** Variation of $2\Delta_0/k_B T_c$ as a function of $T_c$ for NbN films. The *inset* shows the variation of $\Delta_0$ with $T_c$.

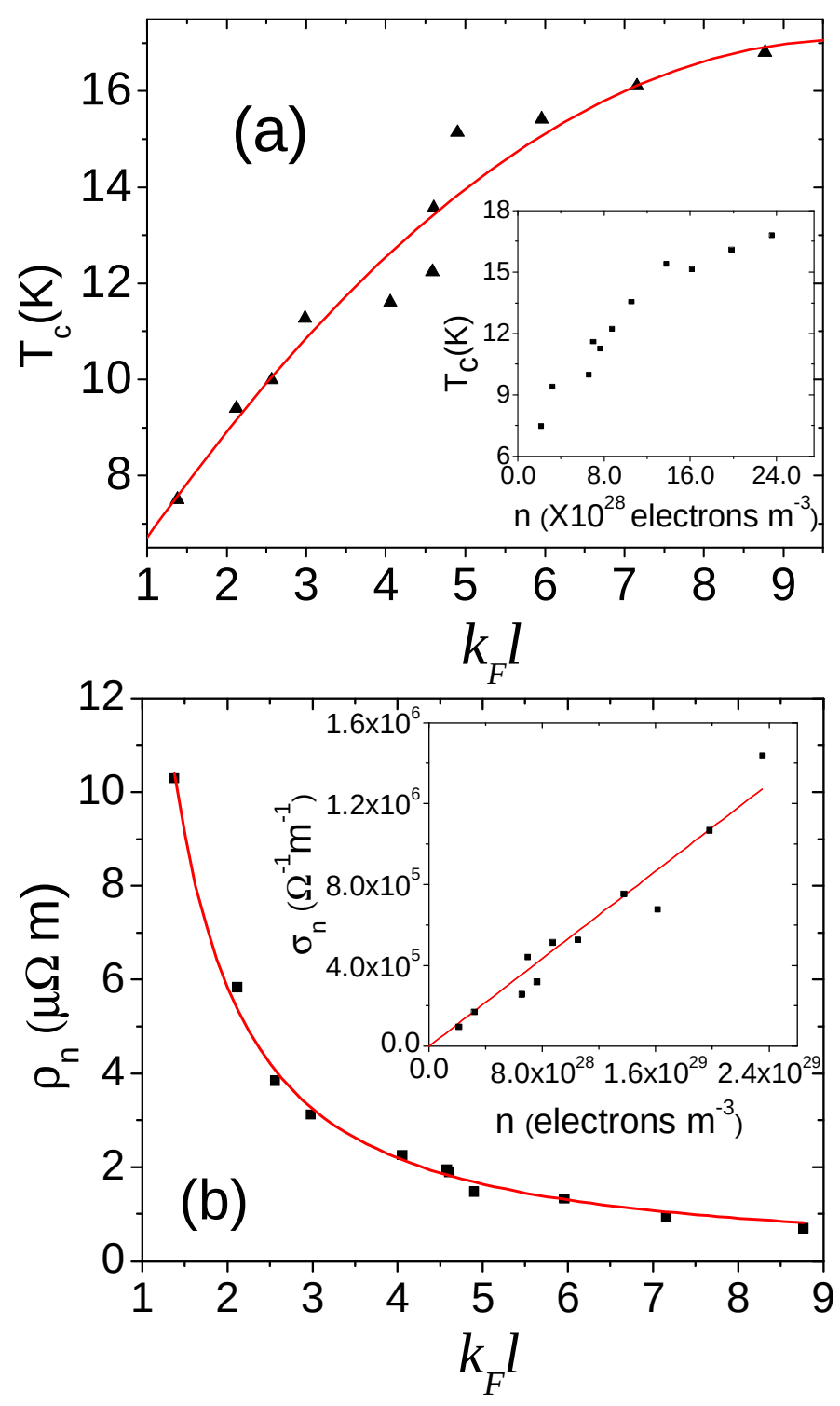

**Figure 1**

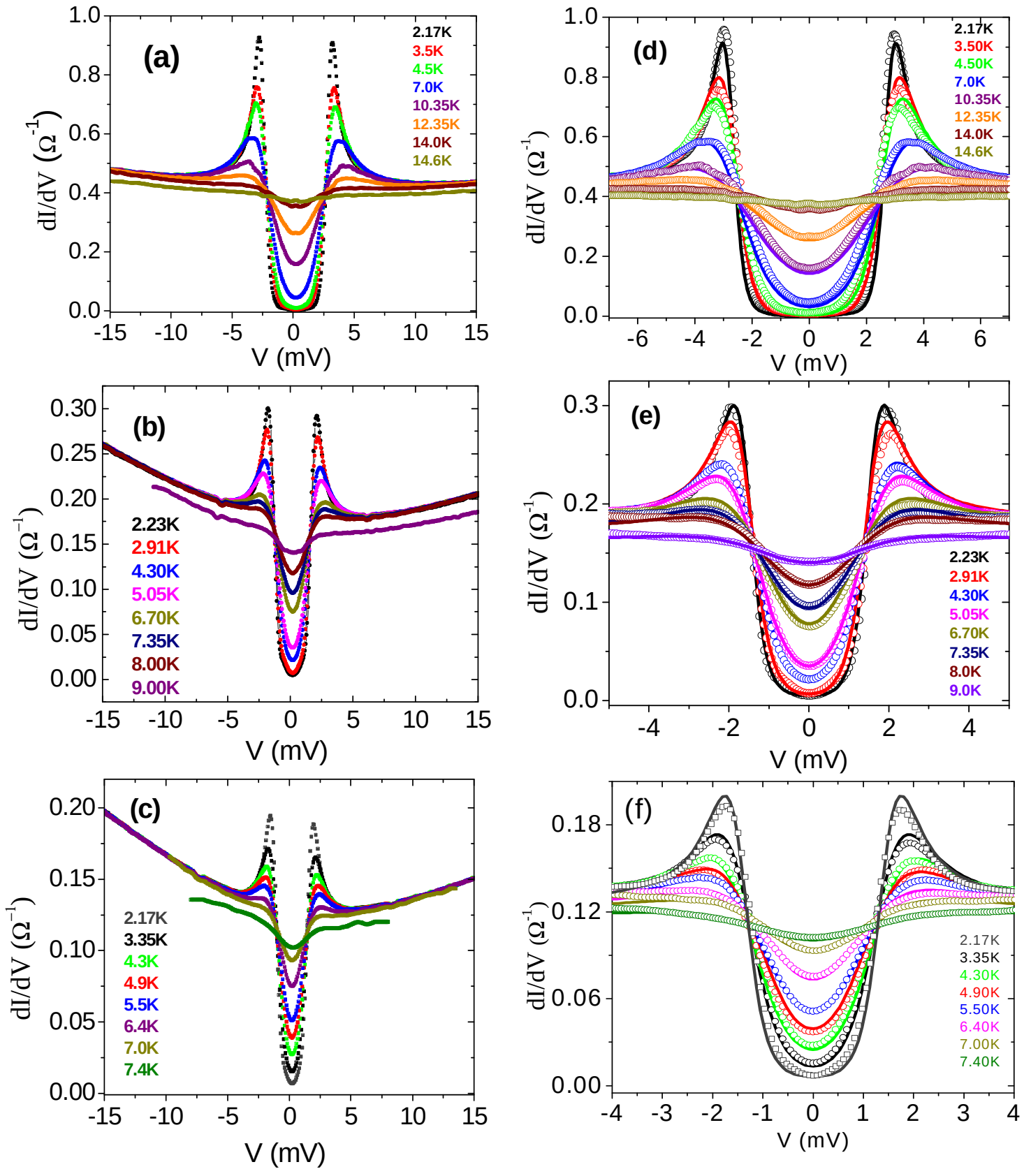

**Figure 2**

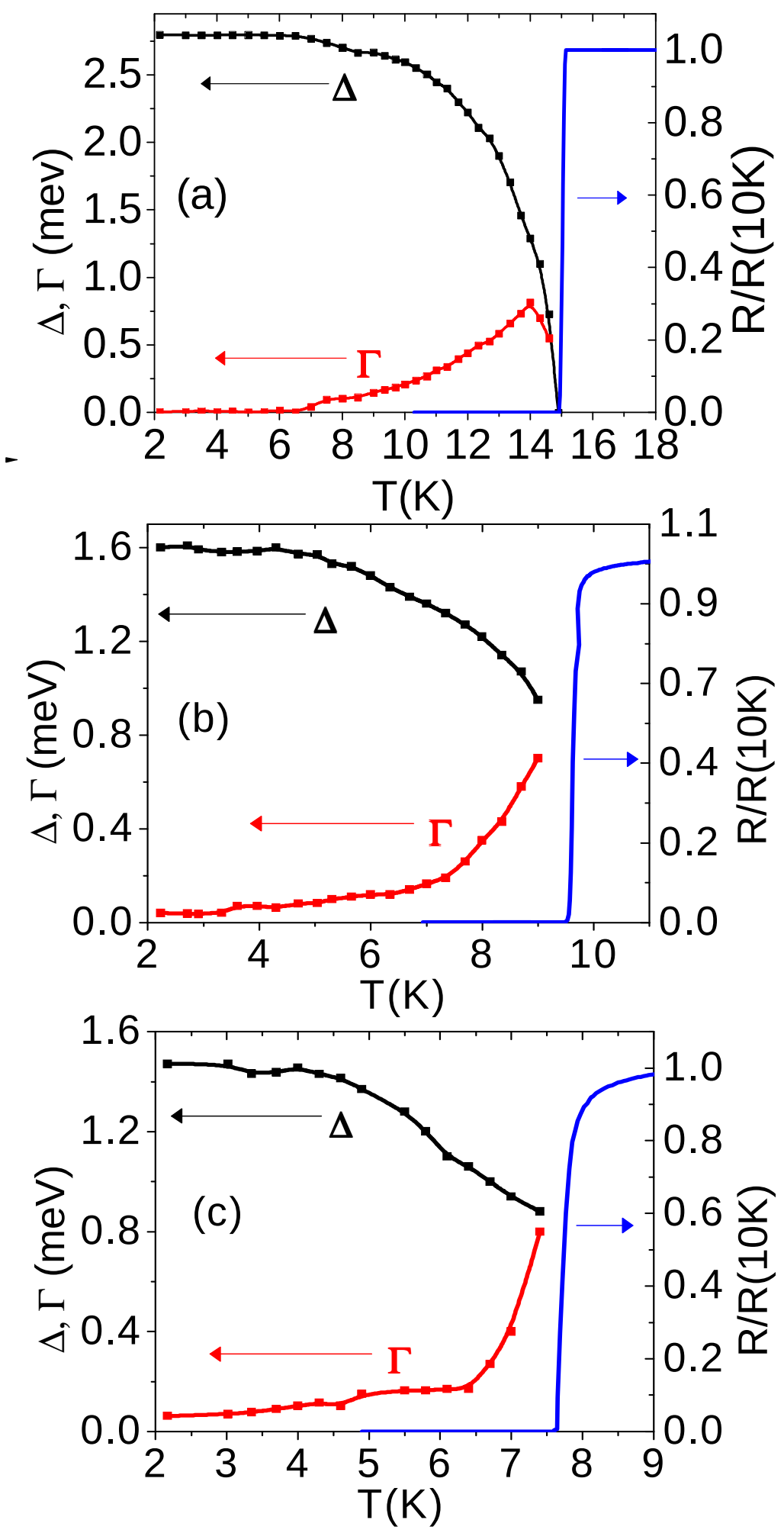

**Figure 3**

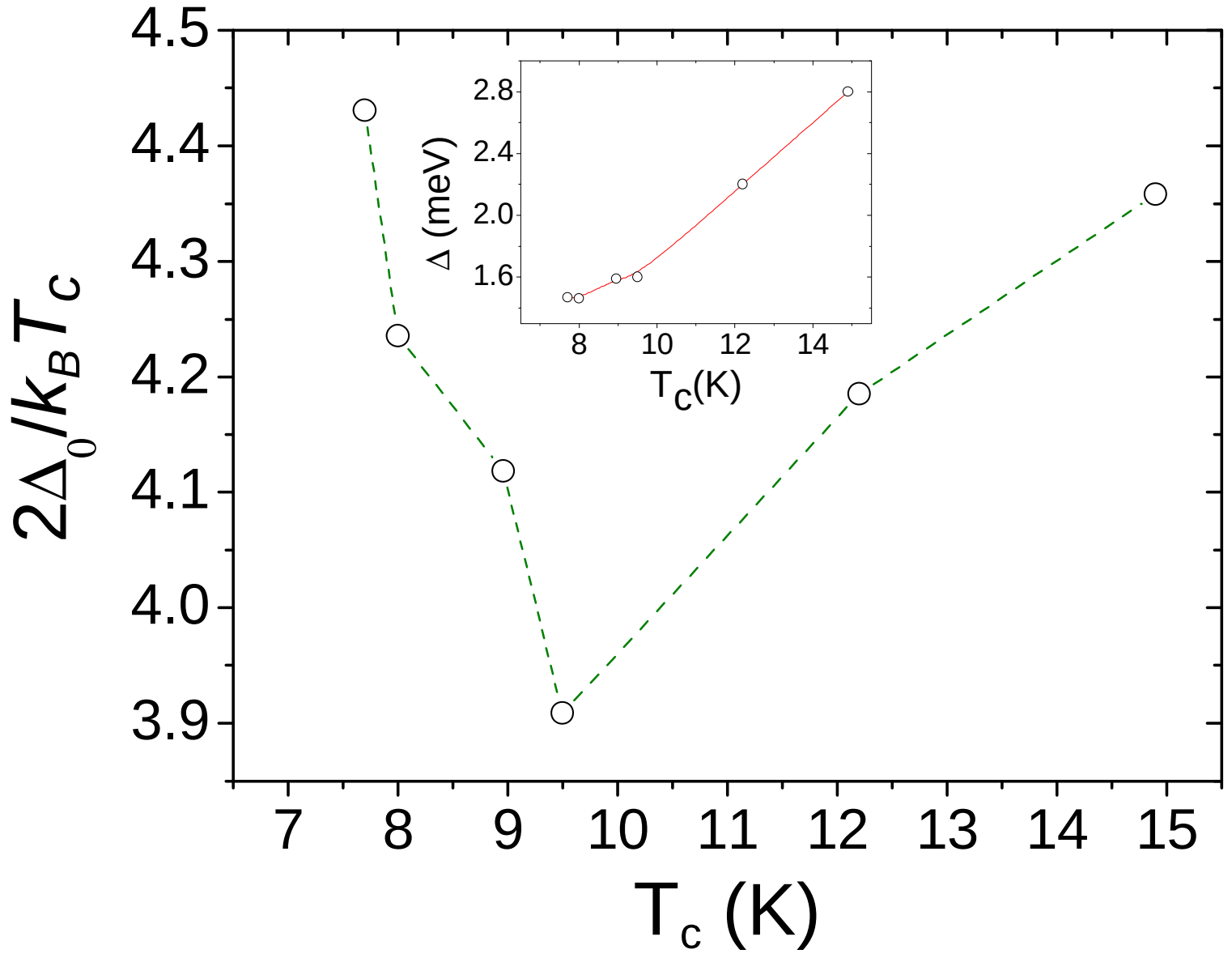
4.5
4.4
4.3
4.2
4.1
4.0
3.9
$2\Delta_0/k_BT_c$
7 8 9 10 11 12 13 14 15
$T_c$ (K)
2.8
2.4
2.0
1.6
Δ (meV)
8 10 12 14
$T_c$(K)

**Figure 4**

[12] Ch. Renner, B. Revaz, J.-Y. Genoud, K. Kadowaki, and Ø. Fischer, Phys. Rev. Lett. **80,** 149 (1998).

[13] L Shan, Y. L. Wang, Y. Huang, S. L. Li, J. Zhao, Pengcheng Dai and H. H. Wen, Phys. Rev. B **78,** 014505 (2008); J. W. Alldredge, J. Lee, K. Mcelroy, M. Wang, K. Fujita, Y. Kohsaka, C. Taylor, H. Eisaki, S. Uchida, P. J. Hirschfeld and J. C. Davis, Nature Physics **4**, 319 (2008); M. Aprili, E. Badica, and L. H. Greene, Phys. Rev Lett. **83,** 4630 (1999).

[14] Mohit Randeria, Rajdeep Sensarma, Nandini Trivedi, and Fu-Chun Zhang, Phys. Rev. Lett. **95,** 137001 (2005).

[15] This symmetrisation procedure did not alter the zero-bias conductance values.

[16] Tinkham, M. *Introduction to Superconductivity* (McGraw-Hill, New York, 1996).

[17] R. C. Dynes, V. Narayanamurti, and J. P. Garno, Phys. Rev. Lett. **41,** 1509 (1978).

[18] The increase in $\Gamma$ as the temperature approaches $T_c$ is expected for a strong coupling superconductor due to phonon mediated recombination of electron and hole like superconducting quasiparticles. See, S. B. Kaplan, C. C. Chi, D. N. Langenberg, J. J. Chang, S. Jafarey, and D. J. Scalapino, Phys. Rev. B **14**, 4854 (1976); R. C. Dynes, V. Narayanamurti, and J. P. Garno, Phys. Rev. Lett. **41**, 1509 (1978).

[19] G. M. Eliashberg, Sov. Phys. JETP **11**, 696 (1960).

[20] (a) T. V. Ramakrishnan, Physica Scripta. **T27,** 24 (1989); (b) K. B. Efetov, Sov. Phys. JETP **51,** 1015 (1980).

[21] Y. Dubi, Y. Meir, and Y. Avishai, Nature (London) **449,** 876 (2007); Y. Dubi, Y. Meir, and Y. Avishai, Physical Review B **78**, 024502 (2008).

[22] The situation is however more complicated than for Josephson Junction network where the amplitude of the order parameter can be reasonably assumed to be temperature independent (see

20). In this case both the amplitude and phase fluctuation have to be treated in a self consistent way.

[23] While we cannot make a detailed comparison with numerical calculations which are performed for T=0, the broadening of the quasiparticle DOS is consistent with simulations. With increase in disorder at T=0, the gap in the quasiparticle spectrum reduces slowly. The BCS DOS on the other hand broadens significantly leading to a complete loss of the BCS coherence peaks at $\pm\Delta$ at high disorder. See, ref. 5.

[24] P. W. Anderson, K. A. Muttalib, and T. V. Ramakrishnan, Phys. Rev. B **28**, 117 (1983).

[25] Leavens has argues that the Anderson-Muttalib-Ramakrishnan (ref. 24) effect is of undetermined magnitude due to the large uncertainty in the critical resistivity; see C. R. Leavens, Phys. Rev. B **31**, 6072 (1985).

[26] G. Bergmann, Z. Physik **228,** 25 (1969).

[27] G. Ziemba and G. Bergmann, Z. Physik **237,** 410 (1970).

[28] M. V. Feigel'man, L. B. Ioffe, V. E. Kravtsov, and E. A. Yuzbashyan, Phys. Rev. Lett. **98,** 027001 (2007).